\documentclass[seceq,preprint]{ptptex}

\let\mbf\boldsymbol
\def\slash#1{\ooalign{\hfil/\hfil\crcr$#1$}}

\def\XZ{{\mbf Z}}
\def\XP{{\mbf P}}
\def\XQ{{\mbf Q}}
\def\XM{{\mbf M}}

\def\XD{{\mbf D}}
\def\XS{{\mbf S}}
\def\XK{{\mbf K}}

\def\XU{{\mbf U}}
\def\XX{{\mbf X}}
\def\XG{{\mbf G}}
\def\XL{{\mbf L}}
\def\XV{{\mbf V}}
\def\XW{{\mbf W}}

\def\dt{\!\cdot\!}
\def\nn{\nonumber\\}

\def\calL{{\cal L}}
\def\calA{{\cal A}}
\def\calD{{\cal D}}
\def\calN{{\cal N}}
\def\hatD{{\hat\calD}}

\def\slashD{{\hat{\slash\calD}}}

\def\hatR{{\hat R}}

\def\calF{{\cal F}}

\def\half{\hbox{\large ${1\over 2}$}}
\def\myfrac#1#2{\hbox{\large ${#1\over #2}$}}
\def\T{{\rm T}}

\def\6#1{{\underline{#1}}}
\def\m6#1{{\underline{#1}\,}}

\newdimen\Tdim
\def\ispan{{\setbox0=\hbox{i}%
\Tdim\ht0\advance\Tdim\dp0\rule[-\dp0]{0pt}{\Tdim}}}
\def\jspan{{\setbox0=\hbox{j}%
\Tdim\ht0\advance\Tdim\dp0\rule[-\dp0]{0pt}{\Tdim}}}
\def\Tspan#1{{\setbox0=\hbox{#1}%
\Tdim\ht0\advance\Tdim\dp0\advance\Tdim.55ex\rule[-\dp0]{0pt}{\Tdim}\box0}}
\def\Span#1{\setbox0=\hbox{$#1$}\rule[-\dp0]{0pt}{\ht0}}

\def\cN{\mathcal{N}}
\def\cA{\mathcal{A}}
\def\cL{\mathcal{L}}
\DeclareMathOperator{\tr}{tr}

\def\vev#1{\langle#1\rangle}

\preprintnumber[4cm]{
\hbox{}\hfill TIT/HEP-562\\ 
\hbox{}\hfill hep-th/0611329}

\title{
Supersymmetric Completion of an $R^2$ Term in Five-Dimensional
Supergravity
}

\author{
 Kentaro 
{\sc Hanaki}$^1$,
Keisuke {\sc Ohashi}$^1$
and Yuji {\sc Tachikawa}$^2$
}

\inst{
$^1$ Department of Physics, Tokyo Institute of Technology, \\
Tokyo 152-8551,  Japan 
\\
$^2$ School of Natural Sciences, Institute for Advanced Study,\\
 Princeton,  New Jersey 08540, USA
}

\abst{
We analyze the structure of a particular
higher derivative correction 
of five-dimensional 
ungauged and gauged supergravity with eight supercharges.
Specifically, 
we determine all the purely bosonic terms which are connected by
the supersymmetry transformation to the mixed gauge-gravitational Chern-Simons 
term, $W \wedge \tr R\wedge R$. Our construction utilizes the superconformal 
formulation of supergravity.
As an application, we determine the condition for the
supersymmetric anti-de Sitter vacuum with this term.
We check that it gives precisely the same condition as
the $a$-maximization in four-dimensional
 superconformal field theory on the boundary,
as predicted by the AdS/CFT correspondence.
}

\begin{document}
\maketitle
\section{Introduction}
Five-dimensional (5d) supergravity with the minimal number\footnote{
Minimal supersymmetry in five dimensions has eight supercharges.
This form of supersymmetry 
is usually called $\cN=2$ in the supergravity literature
and $\cN=1$ in the field theory literature.  We call it $\cN=2$, following
the supergravity convention.}
of supersymmetries, which was first presented in
Refs.\citen{GST1} and \citen{GST2}{},
has applications in our attempt to obtain an understanding of the fundamental properties of nature.
The theory itself and the theory coupled
to a four-dimensional brane are  natural starting points of studies
if one wants to consider a supersymmetric model with
one large extra dimension.  It can also be used for studying
the superconformal field theory in four dimensions,
via the celebrated anti-de Sitter/conformal field theory (AdS/CFT)
correspondence. Also, recently, 
rich exact solutions of the theory, including black rings,
have been obtained. 
Thus, it has been found that this theory 
can be used as a testbed for the dynamical study of gravity.

The Lagrangian of 5d supergravity, however, cannot be thought of
as giving ultraviolet-complete definition of the theory,
since it is not renormalizable. Thus it must be embedded in a 
consistent theory of quantum gravity, say a suitable compactification
of string theory, and the supergravity Lagrangian
should be regarded as an effective description of the low energy limit of
such theories.  Therefore, the Lagrangian should contain
various higher derivative terms, such as a curvature-squared term, $R^2$,
with a small coefficient.  Thus, it is of utmost importance to determine
how  local supersymmetry  governs the higher derivative terms.

Similar problems have been previously studied in six-dimensional \cite{BR} and 
four-dimensional cases.
(See the excellent review in Ref.\citen{MoReview} and references therein.)
In those works,
the superconformal formalism for supergravity was used to facilitate the
analysis. This was useful because the formalism is fully off-shell,
and thus the analysis of the higher derivative terms 
can be done without modifying the supersymmetry transformation laws.

The first objective of this paper is to construct 
a supersymmetric $R^2$ term in 5d $\cN=2$
supergravity using the superconformal formalism
developed in Refs.\citen{KO,FO,Berg0,Berg1}.
More specifically, we study the supersymmetric completion of 
a distinctive class of higher derivative terms
in five dimensions. This class is represented 
by the mixed gravitational Chern-Simons term 
\begin{equation}
W \wedge \tr R\wedge R,
\end{equation} where $W$ is a $U(1)$ gauge field and $R$ is the 
curvature two-form constructed from the metric. 
While we spell out how one can determine all of the terms in the 
supersymmetric completion, we explicitly determine only the purely 
bosonic terms.
The terms we obtain should suffice in the study of purely bosonic backgrounds
and their properties under supersymmetry transformations. Thus we believe
that our results will be of great use.

Although we have not been able to prove that it is the only possible form of 
the four-derivative correction,
we believe the uniqueness of the result, judging from the
structure of the formalism.
Some of the bosonic terms in the completion were known before
using the compactification of the $R^4$ term in M-theory
down to five dimensions\cite{Ferrara1,Ferrara2}. Our result
is consistent with theirs.

As an application,  we study
how the condition for the maximally supersymmetric AdS solution
is modified by the completed $R^2$ term. We see that
it can be mapped to the $a$-maximization of the conformal
field theory on the boundary.

This paper is  organized as follows. 
In \S\ref{sec:review}, we give a brief introduction
to the superconformal formalism in five dimensions.
In \S\ref{sec:construction}, we construct the 
supersymmetric completion of the $W\wedge \tr R\wedge R$
term using the material reviewed in \S\ref{sec:review}.
Then, the analysis of the AdS solution with the
$R^2$ term  is presented in \S\ref{sec:sugraAdS}. 
Finally, we compare the result to the $a$-maximization
of the boundary theory in \S\ref{sec:amax}.
Section \ref{sec:conclusion} gives a summary and discussion.

In this paper, we
mostly
follow the convention of Ref.\citen{FO}{}. 
One difference is
that we explicitly impose constraints to express dependent gauge fields
in terms of composites of independent ones.
This greatly simplifies
the supersymmetry transformation laws and the expression
for the supercovariant curvatures.  
In the hope of making this paper accessible to all readers,
we provide  appendices detailing the notation
and definitions:  
Appendix \ref{sec:notation} contains our conventions, 
Appendix \ref{sec:weylformulae} collects definitions and 
useful formulae 
regarding the Weyl multiplet, and
Appendix \ref{sec:vector-convention} compares various conventions
for vector multiplets in the literature.

\section{Brief review of the superconformal formalism}\label{sec:review}

Ordinary supergravity theories are invariant under 
general coordinate transformations, local Lorentz transformations
and local supertranslations. Thus, they are, in a sense, the gauge theory
of the super-Poincar\'e group. 
For this reason, they are called Poincar\'e supergravities.
The construction of such theories was
originally done by following the Noether procedure,
which yields on-shell actions.

The superconformal approach to Poincar\'e supergravity 
starts from the  construction of theories which are gauge invariant under a much
larger group, the superconformal group in respective dimensions.
Then, by imposing constraints, these theories are identified as gravitational 
theories, so-called conformal supergravities.
The enlargement of the local symmetry greatly facilitates
the determination of the multiplet 
structure and the construction of the invariant action.
As we are not interested in conformal supergravities themselves, 
we arrange one of the scalar fields to take a non-zero vacuum
expectation value (VEV), spontaneously
breaking the conformal supergravity down to
the Poincar\'e supergravity.

This off-shell approach is particularly suited to the construction of the 
higher derivative terms in the supergravity theories. This is because
the superconformal multiplet contains the auxiliary fields, and hence 
the supersymmetry transformation law is independent of the action.
To find a higher derivative term in the on-shell approach,
one needs to consider the modification of the action and of the supersymmetry
transformation  simultaneously. This makes it quite difficult to carry out. 
We see below that having a local superconformal invariance
makes the analysis of the symmetry of the AdS background
surprisingly transparent.

The four main ingredients required to write down the action are the
following:
\begin{enumerate}
\item the structures of various superconformal multiplets,
\item the ``embedding formulae'' which create new multiplets from
      existing multiplets,
\item the ``invariant action formulae'' to form the Lagrangian density,
\item and the gauge fixing down to the Poincar\'e supergravity.
\end{enumerate}
We review each of these in turn. 

\subsection{Superconformal multiplets}
The superconformal algebra relevant for 5d $\cN=2$ supergravity
is the supergroup $F(4)$, with the generators 
\begin{equation}
\XP_a,\quad \XQ_i,\quad \XM_{ab},\quad \XD,\quad 
\XU_{ij},\quad \XS^i,\quad \XK_a,
\end{equation}
where $a,b,\ldots$ are Lorentz indices, and $i,j,\ldots(=1,2)$ are for
the $SU(2)$ doublets.
Here we have suppressed the spinor indices.
The operators $\XP_a$ and $\XM_{ab}$ are the usual Poincar\'e
generators, $\XD$ is the
dilatation, $\XU_{ij}$ is the $SU(2)$ generator,
 $\XK_a$ represents special conformal boosts, $\XQ_i$ is
the $\cN=2$ supersymmetry, and $\XS_i$ is the conformal
supersymmetry.  The charge of the field with respect to the dilatation $\XD$
is called its Weyl weight.
We introduce the gauge fields 
\begin{equation}
e_\mu {}^a,\quad \psi _\mu ^i,\quad \omega _\mu {}^{ab},
\quad b_\mu ,\quad V_\mu ^{ij},\quad \phi _\mu ^i,\quad f_\mu {}^a,
\end{equation}
corresponding, respectively, to the  generators above 
where $\mu ,\nu ,\ldots$ are the world vector indices 
and $\psi _\mu ^i$ and $\phi _\mu ^i$ are $SU(2)$-Majorana
spinors.  The definitions of the covariant derivatives and curvatures are 
given in Appendix \ref{sec:weylformulae}. We first write down 
Yang-Mills theory for this gauge group $F(4)$. Up to this point,
the generators $\XP_a$ and $\XM_{ab}$ have not been 
related to the diffeomorphism, and the transformation
law for various gauge fields follows from the structure constants
of $F(4)$.  

Next, we impose the so-called conventional constraints to identify the 
generators $\XP_a$ and $\XM_{ab}$ as those of the general coordinate 
transformation and the local Lorentz transformation:
\begin{equation}
\hat R_{\mu\nu}{}^a(P)=0,\qquad
\gamma^\mu \hat R^i_{\mu\nu}(Q)=0,\qquad
\hat R_{\mu}{}^a(M)=0\label{conventional-constraints}
\end{equation} 
where the hat denotes supercovariantization, 
and the curvature with respect to a generator ${\bf X}_A$ is
written as $\hat R_{\mu\nu}{}^A(X)$.
These allow us to express the $\XM$,  $\XS$ and $\XK$ gauge fields 
$\omega_\mu{}^{ab}$, $\phi^i_\mu$ 
and $f_\mu{}^a$ in terms of  composite fields 
constructed out of other gauge 
fields.\footnote{As discussed in Ref.\citen{FO}, 
the second and third constraints in (\ref{conventional-constraints}) 
are avoidable and we can keep the gauge fields $f_\mu^a$ and
$\phi^i_\mu$ independent. Here, we impose the constraints to obtain 
simpler transformation laws.}
The local $\XP$, $\XQ$ transformation law needs to be modified
to preserve the constraints \eqref{conventional-constraints}, after which
$e_\mu{}^a$ and $\omega_{\mu}{}^{ab}$ can be identified
with the usual f\"unfbein and the spin connection, respectively.
The $\{\XQ,\XQ\}$ commutator is also modified from that of $F(4)$,
and it is presented below in \eqref{eq:QQcommutator}.
As argued above,  the $\XP$ transformation becomes essentially 
the general coordinate 
transformation $\delta _{\rm GC}(\xi ^\lambda )$:
\begin{eqnarray}
\delta _P(\xi ^a)&=&\delta _{\rm GC}(\xi ^\lambda )-\delta _A(\xi ^\lambda h_\lambda ^A).\label{eq:CGC}
\end{eqnarray}
On a covariant quantity $\Phi $ with only flat indices, 
$\delta _P(\xi ^a)$ acts as  the full covariant derivative:
\begin{equation}
\delta _P(\xi ^a)\Phi =\xi ^a\left(\partial _a-\delta _A(h_a^A)\right)\Phi \equiv \xi ^a\hatD_a\Phi .
\end{equation} 
It is sometimes called the covariant general coordinate transformation.

Next we summarize the structure of the multiplets we use.
Their properties are listed in Table \ref{table} 
of Appendix \ref{sec:notation}.

\subsubsection{The Weyl multiplet}
We add auxiliary fields $v^{ab}$, $\chi^i$ and $D$ to the set of  
gauge fields above to obtain an 
irreducible Weyl multiplet, which
consists of 32 bosonic plus 32 fermionic component fields,
\begin{eqnarray}
e_\mu {}^a,\quad \psi _\mu ^i,\quad V_\mu ^{ij},\quad b_\mu ,\quad v^{ab},
\quad \chi ^i,\quad D,
\end{eqnarray} 
where $v^{ab}$ is antisymmetric in $a$ and $b$,
$\chi^i$ is an $SU(2)$-Majorana spinor, and $D$ is a scalar.
The  $\XQ, \XS$ and $ \XK$ transformation laws for the Weyl 
multiplet are as follows [with
 $\delta \equiv \bar\varepsilon ^i\XQ_i+\bar\eta ^i\XS_i+\xi _K^a\XK_a\equiv \delta _Q(\varepsilon )+
\delta _S(\eta )+\delta _K(\xi _K^a)$]:
\begin{eqnarray}
\delta e_\mu {}^a&=&-2i\bar\varepsilon \gamma ^a\psi _\mu ,\nn
\delta \psi _\mu ^i&=&{\cal D}_\mu \varepsilon ^i+\myfrac12 v^{ab}\gamma _{\mu ab}\varepsilon ^i-\gamma _\mu \eta ^i,
\nn
\delta b_\mu &=&-2i\bar\varepsilon \phi _\mu -2i\bar\eta \psi _\mu -2\xi _{K\mu },\nn
\delta V_\mu ^{ij}&=&-6i\bar\varepsilon ^{(i}\phi ^{j)}_\mu 
+4i\bar\varepsilon ^{(i}\gamma \dt v\psi ^{j)}_\mu 
-\myfrac{i}4\bar\varepsilon ^{(i}\gamma _\mu \chi ^{j)}+6i\bar\eta ^{(i}\psi ^{j)}_\mu ,\nn
\delta v_{ab}
&=&-\myfrac{i}8\bar\varepsilon \gamma _{ab}\chi 
-\myfrac32 i\bar\varepsilon \hatR_{ab}(Q), \nn
\delta \chi ^i&=&D\varepsilon ^i-2\gamma ^c\gamma ^{ab}\varepsilon ^i\hatD_av_{bc}+\gamma \dt \hatR(U)^i{}_j\varepsilon ^j
-2\gamma ^a\varepsilon ^i\epsilon_{abcde}v^{bc}v^{de}+4\gamma \dt v\eta ^i,\nn
\delta D&=&-i\bar\varepsilon \slashD\chi -8i\bar\varepsilon \hatR_{ab}(Q)v^{ab}+i\bar\eta \chi,
\label{eq:susytr-weyl}
\end{eqnarray}
where the derivative $\calD_\mu $ is covariant only with respect to the
homogeneous transformations $\XM_{ab},\XD$ and $\XU^{ij}$. The dot 
product $\gamma \dt T$ for a tensor $T_{ab\cdots  }$ generally represents the 
contraction $\gamma ^{ab\cdots  }T_{ab\cdots  }$. 
When the $SU(2)$ indices are suppressed
in bilinear terms of spinors, the northwest-southeast
 contraction, $\bar\psi\gamma^{a_1\ldots a_n}\lambda=\bar\psi^i\gamma^{a_1\ldots a_n}\lambda_i$, is understood.
The algebra of the $\XQ$ and $\XS$ transformations takes the form
\begin{eqnarray}
{}[\delta _Q(\varepsilon _1),\,\delta _Q(\varepsilon _2)]&=&\delta _P(2i\bar\varepsilon _1\gamma _a\varepsilon _2)
+\delta _M(2i\bar\varepsilon _1\gamma ^{abcd}\varepsilon _2v_{ab})
+\delta _U(-4i\bar\varepsilon _1^i\gamma \dt v\varepsilon _2^j)\nn
&&+\delta _S\left(\cdots\right)
+\delta _K\left(\cdots
\right),\label{eq:QQcommutator}\\
{}[\delta _S(\eta ),\delta _Q(\varepsilon )]&=&\delta _D(-2i\bar\varepsilon \eta )
+\delta _M(2i\bar\varepsilon \gamma ^{ab}\eta )+\delta _U(-6i\bar\varepsilon ^{(i}\eta ^{j)})\nn
&&+\delta _K\left(\cdots
\right),\label{eq:SQcommutator}
\end{eqnarray}
where the translation $\delta _P(\xi ^a)$ is defined in \eqref{eq:CGC}.
We summarize the useful formulae for the supercovariant derivatives and 
curvatures in Appendix \ref{sec:weylformulae}.

One particular point which is relevant in the analysis given in
 subsequent sections
 is that the third constraint in \eqref{conventional-constraints}
makes the supercovariant curvature $\hatR(M)_{abcd}$
traceless. Thus, for a background in which the 
nontrivial component of the Weyl multiplet is only the f\"unfbein, 
$\hatR_{\mu\nu}^{ab}(M)$ is the Weyl tensor of the metric, i.e.
\begin{equation}
\hatR_{\mu\nu}{}^{ab}(M)=R_{\mu\nu}{}^{ab}
+\myfrac43 R^{[a}_{[\mu} e^{b]}_{\nu]}
-\myfrac16 e^{[a}_{[\mu} e^{b]}_{\nu]}R,
\label{eq:R-is-traceless}
\end{equation}where $R_{abcd}$ is the ordinary curvature tensor 
constructed from the metric.

\subsubsection{Vector multiplet}
The vector multiplet consists of 
\begin{equation}
W_\mu^I,\qquad M^I, \qquad \Omega_i^I,\qquad Y_{ij}^I,
\end{equation}
where the index $I$ labels the generators $T_I$ of
the gauge group $G$. Here, $W_\mu^I$ are the gauge fields, $M^I$ are 
the scalar fields
in the vector multiplet, $\Omega^I$ are the 
$SU(2)$-Majorana gaugini, and $Y^I_{ij}$
are  $SU(2)$-triplet auxiliary fields.
We set $W_\mu \equiv W^I_\mu T_I$, and
similarly for other components.
The $\XQ$ and $\XS$ transformation laws of the vector multiplet are then given by
\begin{eqnarray}
\delta W_\mu &=&-2i\bar\varepsilon \gamma _\mu \Omega +2i\bar\varepsilon \psi _\mu M,\nn
\delta M&=&2i\bar\varepsilon \Omega ,\nn
\delta \Omega ^i&=&-\myfrac14\gamma \dt \hat F(W)\varepsilon ^i
-\myfrac12\slashD M \varepsilon ^i+Y^i{}_j\varepsilon ^j-M\eta ^i,\nn
\delta Y^{ij}&=&2i\bar\varepsilon ^{(i}\slashD\Omega ^{j)}-i\bar\varepsilon ^{(i}\gamma \dt v\Omega ^{j)}
-\myfrac{i}4\bar\varepsilon ^{(i}\chi ^{j)}M
-2ig\bar\varepsilon ^{(i}[M,\Omega ^{j)}]-2i\bar\eta ^{(i}\Omega ^{j)}.
\label{eq:susytr-vec}
\end{eqnarray}The transformation law of the gauge
field $W_\mu $ above shows that the superconformal group and 
the gauge group $G$ are not separate but 
have non-zero structure functions, 
$f_{PQ}{}^G$ and $f_{QQ}{}^G$, between them.
For consistency, it is thus required that
 the commutator of two $\XQ$ transformations is modified to 
\begin{equation}
[\delta _Q(\varepsilon _1),\,\delta _Q(\varepsilon _2)]=({\rm R.H.S.~of~(\ref{eq:QQcommutator})})
+\delta _G(-2i\bar\varepsilon _1\varepsilon _2M).\label{eq:QQcommutator2}
\end{equation}
Thus, the supercovariant curvature is given by 
\begin{equation}
\hat F_{\mu\nu}(W)=2\partial_{[\mu}W_{\nu]}-g[W_\mu,W_\nu]
+4i\bar\psi_{[\mu}\gamma_{\nu]}\Omega-2i\bar\psi_\mu\psi_\nu M.
\end{equation}

\subsubsection{Hypermultiplet}
The hypermultiplet in 5D consists of scalars $\calA^i_\alpha $,
spinors $\zeta _\alpha $ and auxiliary fields $\calF^i_\alpha $. They carry the
index $\alpha ~(=1,2,\ldots,2r)$ of $USp(2r)$.
The scalars satisfy the reality condition $\cA^i_\alpha=-(\cA_i^\alpha)^*$,
and the spinors $\zeta _\alpha $  are $USp(2r)$-Majorana.
 A subgroup $G'$ of the gauge group $G$ 
can act on the index $\alpha$ as a subgroup of $USp(2r)$.
The $\XQ$ and $\XS$ transformations of
 $\calA^i_\alpha $ and $\zeta _\alpha $
are given by
\begin{eqnarray}
\delta \calA^i_\alpha &=&2i\bar\varepsilon ^i\zeta _\alpha ,\nn
\delta \zeta ^\alpha &=&\slashD\calA^\alpha _j\varepsilon ^j-\gamma \dt v\varepsilon ^j\calA^\alpha _j
-gM_*\calA^\alpha _j\varepsilon ^j+3\calA^\alpha _j\eta ^j,
\end{eqnarray} 
where $\slashD$ and $M_*$
include the `central charge' gauge transformation $\XZ$. The quantity $g$ is the coupling constant,
and the notation $X_*Y$ represents generator of the gauge transformation,
\begin{equation}
(X_* Y)^\alpha=X^I t_I{}^\alpha{}_\beta Y^\beta+X^{0}{\XZ}Y^\alpha,
\end{equation} where $X$ takes values in a Lie algebra,
$Y$ takes values in its representation, and 
$t_I{}^\alpha{}_\beta$ is the representation matrix. 
The closure of the algebra thus determines the `central charge'
gauge transformation of $\calA_\alpha^i$ via $\calF^i_\alpha$,
though we set ${\XZ}=0, (\calF^i_\alpha=0)$ in this paper.
(Interested readers are referred to Refs.\citen{KO}and \citen{FO} for details.)

\subsubsection{Linear Multiplet}
A linear multiplet consists of 
\begin{equation}
L^{ij},\quad \varphi^i,\quad E_a,\quad N,
\end{equation} 
where $L^{ij}$ is symmetric in $i$ and $j$ and is real,
$\varphi^i$ is $SU(2)$-Majorana, $E_a$ is a vector, and $N$ is a scalar.

The $\XQ$ and $\XS$ transformation laws of the linear multiplet are
given by
\begin{eqnarray}
\delta L^{ij}&=&2i\bar\varepsilon ^{(i}\varphi ^{j)},\nn
\delta \varphi ^i&=&-\slashD L^{ij}\varepsilon _j
+\myfrac12\gamma ^a\varepsilon ^iE_a
+\myfrac12\varepsilon ^iN\nn
&&
+2\gamma \dt v\varepsilon _jL^{ij}
+gM_*L^{ij}\varepsilon _j-6L^{ij}\eta _j
,\nn
\delta E^a&=&2i\bar\varepsilon \gamma ^{ab}\hatD_b\varphi 
-2i\bar\varepsilon \gamma ^{abc}\varphi v_{bc}
+6i\bar\varepsilon \gamma _b\varphi v^{ab}\nn
&&
+2ig\bar\varepsilon \gamma ^aM_*\varphi 
-4ig\bar\varepsilon ^i\gamma ^a\Omega ^j_*L_{ij}
-8i\bar\eta \gamma ^a\varphi, \nn
\delta N&=&-2i\bar\varepsilon \slashD\varphi -3i\bar\varepsilon \gamma \dt v\varphi 
+\myfrac12i\bar\varepsilon ^i\chi ^jL_{ij}
+4ig\bar\varepsilon ^{(i}\Omega ^{j)}_*L_{ij}
-6i\bar\eta \varphi. \label{eq:trf.L}
\end{eqnarray}
The algebra closes if $E^a$ satisfies the following $\XQ$- and
$\XS$-invariant constraint:
\begin{equation}
\hatD_aE^a +gM_*N+4ig\bar\Omega_* \varphi+2gY^{ij}_*L_{ij}
=0.
\label{eq:Con.E}
\end{equation}

An important property concerning the linear multiplet is that 
any symmetric, real composite bosonic field
$L^{ij}$, which is
invariant under $\XS$ transformations, automatically
leads to the above transformation law with
suitable choices of $\varphi^i$, $E^a$ and $N$.
Thus,  the construction of a linear multiplet can be carried out
by repeated supersymmetric transformations starting from
the lowest component, $L^{ij}$.

\subsection{Embedding formulae}
Vector multiplets can be embedded into a linear multiplet given an
arbitrary quadratic homogeneous polynomial $f(M)$ 
of the first components $M^I$
of the vector multiplets.  
They are given by
\begin{eqnarray}
L_{ij}(\XV )&=& Y_{ij}^If_I-i\bar\Omega ^I_i\Omega ^J_jf_{IJ},\nn
\varphi _i(\XV )&=&-\myfrac14\chi _if\nn
&&+\left(\slashD\Omega ^I_i-\myfrac12\gamma \dt v\Omega _i^I-g[M,\Omega ]^I\right)f_I\nn
&&+\left(-\myfrac14\gamma \dt\hat F^I(W)\Omega ^J+\myfrac12\slashD M^I\Omega ^J
-Y^I\Omega ^J\right)f_{IJ},\nn
E_a(\XV )&=&\hatD^b\left(4v_{ab}f+\hat F_{ab}^I(W)f_I
+i\bar\Omega ^I\gamma _{ab}\Omega ^Jf_{IJ}\right)\nn
&&+\left(-2ig[\bar\Omega ,\gamma _a\Omega ]^I+g[M,\hatD_aM]^I
\right)f_I\nn
&&+\left(-2ig\bar\Omega ^I\gamma _a[M,\Omega ]^J+\myfrac18\epsilon _{abcde}\hat
F^{bcI}(W)\hat F^{deJ}(W)\right)f_{IJ},\nn
N(\XV )&=&-\hatD^a\hatD_af
+\left(-\myfrac12 D -3v^2\right)f\nn
&&+\left(-2\hat F_{ab}(W)v^{ab}+i\bar\chi \Omega ^I+2ig[\bar\Omega ,\Omega ]^I\right)f_I\nn
&&+\left(\begin{array}{c}
         -\myfrac14\hat F_{ab}^I(W)\hat F^{abJ}(W) 
         +\myfrac12\hatD^aM^I\hatD_aM^J\\
           +2i\bar\Omega ^I\slashD\Omega ^J-i\bar\Omega ^I\gamma \dt v\Omega ^J+Y^I_{ij}Y^{Jij}
\end{array}
\right)f_{IJ},\label{eq:VtoL}
\end{eqnarray}
where a scalar function with the subscripts $I,J,\ldots$
represent for  its repeated derivative with respect to $M^{I,J,\ldots}$.
For example,  $f_{IJ}\equiv \partial_I \partial_J f$.
We often use the notation $v^2$ for $v_{ab}v^{ab}$ in this paper.
One can also form a linear multiplet from two hypermultiplets.

\subsection{Invariant action formulae}
We can construct an invariant action from a pair of vector and 
linear multiplets as 
\begin{eqnarray}
e^{-1}{\cal L}(\XV\cdot\XL)
&\equiv &Y^{ij}\cdot L_{ij}+2i\bar\Omega \cdot \varphi
 +2i\bar\psi ^a_i\gamma _a\Omega _j\cdot L^{ij}\nn
&&-\myfrac12W_a\cdot \left(E^a-2i\bar\psi _b\gamma ^{ba}\varphi +2i\bar\psi _b^{(i}\gamma 
 ^{abc}\psi _c^{j)}L_{ij}\right)\nn
&&+\myfrac12M\cdot \left(N-2i\bar\psi _b\gamma ^{b}\varphi 
-2i\bar\psi _a^{(i}\gamma ^{ab}\psi _b^{j)}L_{ij}\right),\label{eq:InvAction}
\end{eqnarray} 
where we have restricted our consideration 
to the case that $L^{ij}$ is neutral, for simplicity.

As we have seen, one can form a linear multiplet from two vector
multiplets 
$V^I$ and $V^J$
by using the embedding formula. Then, the invariant action formula above 
can combine it with another vector multiplet $V^K$ to form an action.
The resulting action is 
 completely symmetric in $I$, $J$ and $K$. Thus, we obtain 
an invariant action $\cL_V$ given a gauge-invariant
cubic function $\cN=c_{IJK}M^IM^JM^K$.
For brevity, we consider the case $G=U(1)^{n_v+1}$.
Then the  bosonic term is 
\begin{eqnarray}
e^{-1}{\cal L}_{V}\big|_{\text{bosonic}}&=&
\calN\left(
-\myfrac12D +\myfrac14R(M)
-3v^2\right)+\calN_I\left(
-2v^{ab} F_{ab}^I(W)
\right)\nn
&&{}+\calN_{IJ}\left(
-\myfrac14 F_{ab}^I(W) F^{abJ}(W)+\myfrac12\calD^aM^I\calD_aM^J
+Y_{ij}^IY^{Jij}\right)\nn
&&{}-e^{-1}\myfrac1{24}\epsilon ^{\lambda \mu \nu \rho \sigma }
\cN_{IJK}W_\lambda ^IF_{\mu \nu }^J(W)F_{\rho \sigma }^K(W).\label{eq:vector-action}
\end{eqnarray}
Note the appearance of the gauge Chern-Simons interaction, $W\wedge F\wedge F$,
which came from the $W_a\cdot E^a$ term in the invariant action formula.
The strength of the Chern-Simons interaction directly gives the function $\cN$.
Thus, it governs the entire vector-multiplet Lagrangian.

For the hypermultiplets, the combination of the embedding into the linear multiplet
and the $V\cdot L$ action formula gives an action with
the following bosonic terms: 
\begin{eqnarray}
e^{-1}{\cal L}_H\big|_{\text{bosonic}}&=&
\calD^a\calA^{\bar\alpha }_i\calD_a\calA_\alpha ^i
+\calA^{\bar\alpha }_i(gM)^2\calA^i_\alpha \nn
&&{}+\calA^2\left(\myfrac18D 
+\myfrac3{16}R(M) -\myfrac14v^2\right)
+2gY^{ij}_{\alpha\beta}\calA^{\bar\alpha }_i\calA_j^\beta, 
\label{eq:hyper-action}
\end{eqnarray} 
where $\cA^2\equiv \cA^{\bar \alpha}_i\cA_\alpha^i
=\cA^{\beta}_id_\beta{}^\alpha\cA_\alpha^i $ with the metric 
$d_\alpha{}^\beta$ arranged to be $\delta_\beta{}^\alpha$ for a compensator.
We have already eliminated the auxiliary fields $\calF_{i\alpha}$
using their equations of motion.

\subsection{Gauged supergravity}
Let us now consider a system coupled to $n_V+1$ 
conformal vector multiplets, $I=0,\ldots, n_V$,
and one conformal hypermultiplet, $\cA^i_\alpha$ ($i,\alpha=1,2$), 
as a compensator. We let its action be $\cL_0=\cL_H -\myfrac12\cL_V$.
The equation of motion for $D$ gives $\cA^2+2\cN=0$,
while the scalar curvature appears in the Lagrangian
in the form
\begin{equation}
(\myfrac3{16}\cA^2-\myfrac18\cN) R(M).
\end{equation} 
Thus, we can make the Einstein-Hilbert term
canonical by fixing the dilatational gauge transformation $\XD$
via the condition $\cA^2=-2$. 
It also fixes $\cN=c_{IJK}M^IM^JM^K =1$ via the equation of motion for $D$. 
Thus, the scalars parametrize a `very special' 
manifold.\cite{veryspecial}

It is known that the AdS background requires gauged
supergravity, which is obtained by introducing a charged compensator. 
Therefore, let us consider a model with charges  
$G_I\cA^\alpha_i=P_I(i\sigma^3)^\alpha{}_\beta\cA^\beta_i$ and 
fix the $\XU$-gauge transformation by $\cA^\alpha_i=\delta^\alpha_i$. 
Under this fixing, only the combination 
\begin{eqnarray}
 \delta_{G}'(\Lambda^I)=\delta_{G}(\Lambda^I)
+\delta_U\left(\Lambda^IP_I(i\sigma^3)^{ij}\right)\label{eq:modG}
\end{eqnarray}  
of $U(1)$ gauge and $\XU$ 
transformations survives. In this model, the vectors $W^I_\mu$ are coupled to the hypermultiplet
via\footnote{We usually take the $\XD$ gauge field $b_{\mu}$ to be zero 
through $\XK$ gauge fixing.} 
\begin{equation}
{\cal D}_\mu\cA^\alpha_i= \partial_\mu\cA^\alpha_i- W^I_{\mu} P_I(i \sigma^3)^\alpha{}_\beta\cA^\beta_i+\cA^\alpha_jV_\mu^j{}_i,
\end{equation}
where $V^{ij}_\mu$ is the gauge field for the $\XU$ transformation.
The equation of motion for $V_\mu^{ij}$ 
and the condition $\cA^\alpha_i=\delta^\alpha_i$ require 
\begin{equation}
V^{ij}_\mu=P_I (i \sigma^3)^{ij} W^I_\mu,\label{eq:VWmixing}
\end{equation}
which is preserved under the transformations $\delta_G'(\Lambda)$.  
Thus, any $SU(2)_R$ doublet becomes effectively 
charged with respect to the vectors $W^I_\mu$ through \eqref{eq:VWmixing},
with the charges $P_I$.  The auxiliary fields $v^{ab}$ and
$Y^I_{ij}$ are determined to be
\begin{equation}
v_{ab}=-\cN_I F^I_{ab} / 4\cN,\qquad
Y^I_{ij}=2(\cN^{-1})^{IJ} P_I(i \sigma^3)_{ij},\label{eq:aux-sol}
\end{equation}
where $(\cN^{-1})^{IJ}$ is the inverse of $\cN_{IJ}$.
Then, the scalar potential $V$ is given by
 \begin{equation}
V=-4(\cN^{-1})^{IJ}P_IP_J - 2(P_IM^I)^2.
\end{equation} 
By changing the convention to that employed by 
G\"unaydin, Sierra and Townsend in Refs.\citen{GST1} and \citen{GST2} 
via the dictionary given in
in Appendix \ref{sec:vector-convention}, 
and using various identities of the very special geometry, 
it can be shown that  
\begin{equation}
V=3 g^{xy} \partial_x h^I \partial_y h^J P_I P_J
-4 (P_I h^I)^2,
\end{equation} 
which is the usual form presented in 
the supergravity Lagrangian in the on-shell formalism.

The above procedure reproduces the structure of 5d $\cN=2$
gauged supergravity in the on-shell formalism, as should be the case.
We use the action $\cL_H -\myfrac12\cL_V$ as the zeroth-order term,
where we add the $R^2$ term to be constructed below.

\section{Construction of a supersymmetric $R^2$ term}\label{sec:construction}

\subsection{Strategy}
Before moving on,
we need to make a few comments on the physical interpretation
of the higher derivative terms, in particular in the off-shell formalism.
Firstly,  if we naively apply the variational method to obtain the
equation of motion from a higher derivative theory,
it results in a differential equation which is higher than second order.
This means that giving the value and the first derivative of a field
does not suffice as initial values. In other words, there are `extra modes'
in addition to the modes of the two-derivative Lagrangian.
This is inevitable if we take the Lagrangian as giving 
an ultra-violet definition.

However, we regard our  Lagrangian  to be the effective
low-energy description
in a derivative expansion with a small expansion parameter $\alpha'$.
Thus, the solution to the equation of motion
should take the form of  a perturbative expansion in $\alpha'$, and,
 in particular, its $\alpha'\to 0$ limit should exist. 
Such solutions are known to be determined by
the value and the first derivative of a field at $t=0$, just as
in the case with two-derivative Lagrangian,
making the `extra modes' mentioned above unphysical.
(The details can be found, for example, in 
Refs.\citen{interpretation-of-higher}
and \citen{interpretation-of-higher2}.)

Secondly, it is readily checked that the auxiliary fields would
appear with physical kinetic terms and begin to propagate
when one constructs higher derivative terms in the off-shell formalism.
It is known, however, that the auxiliary fields can be eliminated 
perturbatively in $\alpha'$ (see e.g. the introduction of
Ref.\citen{aux-elimination}) to
produce many higher derivative terms in the physical fields. 
The resulting Lagrangian is to be understood as explained in the previous
paragraph. Thus, the would-be propagating auxiliary fields are just the `extra modes'
associated with the higher derivative terms, and they are not to be regarded
as physical fields.

The third comment is of a slightly different nature. In the higher derivative theory
of gravity, one can redefine the metric as \begin{equation}
g_{\mu\nu}\to g_{\mu\nu} + aRg_{\mu\nu}+b R_{\mu\nu}+\cdots,
\label{eq:redefinition}
\end{equation}with $a$ and $b$ small parameters. 
This leaves the leading-order Einstein-Hilbert term intact, while changing
the form of  the higher-order derivative terms.  
For example, it can be used
to arbitrarily shift the coefficients of $R^{\mu\nu}R_{\mu\nu}$
and $R^2$, while that of $R^{\mu\nu\rho\sigma}R_{\mu\nu\rho\sigma}$ cannot
be shifted.
It should also change
the supersymmetry transformation law.  The physics described
by the Lagrangian, of course, remains the same under the redefinition.
We need to use a redefinition to compare our results
to those in literature.

Below, we construct a very specific higher derivative term,
whose form is not preserved by \eqref{eq:redefinition}.
This is because we use a very specific form of the supersymmetry transformation
dictated by the superconformal formalism.
Change in the conventional constraints \eqref{conventional-constraints}
also induces a field redefinition among the fields in the Weyl multiplets
without altering physical contents of the theory.
Our choice of the constraint $\hatR^a_\mu=0$ is a convenient one
because it greatly reduces the number of higher derivative terms to consider
by forbidding the appearance of the terms like $\hatR_{ab} \hatR^{ab}$
or $\hatR^2$.

With these preliminary remarks,
we set out to construct a supersymmetric curvature-squared term
in 5d $\cN=2$ supergravity.  More precisely, we obtain the supersymmetric
completion of the mixed gauge-gravitational Chern-Simons term, 
\begin{equation}
\epsilon^{abcde}W^I_a  R_{bc}{}_{fg} R_{de}{}^{fg}. \label{eq:gCS}
\end{equation} 
We recall that the gauge Chern-Simons term in \eqref{eq:vector-action}
arises from
the $W_a \cdot E^a $ term   in the $\XV\cdot \XL$ invariant action formula.
Judging from the similarity of the roles played by the gauge curvature $F^I_{ab}$ 
and the metric curvature $R_{ab}{}^{cd}$, a natural guess would be
to first embed the Weyl multiplet into a vector multiplet $V^{cd}[\XW]$ with extra
antisymmetric Lorentz indices $c$ and $d$, and then to construct a linear multiplet
from the $\XL(V^I, V^J)$ embedding formula. 
However, we have found that this method is not significantly
better than the direct construction of the linear multiplet. 
Therefore, our strategy is as follows:
\begin{enumerate}
\item embed the Weyl multiplet to the linear multiplet;
\item use the $\calL(\XV\cdot\XL)$ invariant action formula;
\item gauge-fix down to the Poincar\'e supergravity.
\end{enumerate}

We believe that the combination we determine is the most general
form of  supersymmetric coupling between
the Weyl multiplet  and vector multiplets, 
although we do not have
a definite proof.  The following is a rough argument. 
Suppose that a supersymmetric combination of four-derivative terms
contains a term of the form $f( M,\hatR(M)^2 ) \hatR(M)^2$. 
Then $f(M,\hatR(M)^2)$ 
must be of Weyl weight 1.  If this is not a linear combination
of $M^I$, its supersymmetric completion will contain a term of the form
$ c_I(M,\hatR(M)^2) W^I\wedge \tr \hatR(M)\wedge\hatR(M)$, with non-trivial
function $c_I$ of zero Weyl weight. This term is not gauge invariant
unless $c_I$ is constant. Thus $f(M)$ is necessarily of the form $c_I M^I$
for constant $c_I$.  This is in stark contrast with the situation in 
four-dimensional $\cN=2$ supergravity, where \cite{MoReview} 
one can use an arbitrary holomorphic function $F(X^I, \XW^2)$
constructed from the vector multiplet scalar $X^I$ and the 
scalar $\XW^2$ constructed from the square of the Weyl multiplet.

The strong restriction in five dimensions, of course,
already appears in the two-derivative Lagrangian. Indeed, 
the structure of the four-dimensional $\cN=2$ vector multiplet
is determined by a holomorphic function $F(X^I)$, but
in five dimensions, the corresponding object $\cN$ 
must be a purely cubic function. 
This restriction comes from the gauge invariance
of the gauge Chern-Simons terms, just as in the case considered above.  

\subsection{Embedding and an invariant action}
The linear multiplet should have
$E_a\ni \epsilon _{abcde}R^{bc}{}_{fg}(M)R^{defg}(M)$
to be used in the invariant action formula in order to obtain the 
gravitational Chern-Simons term \eqref{eq:gCS}.
The supertransformation law \eqref{eq:R(Q)tr} for $\hat R(Q)$  reveals 
that we need the following structure:
\begin{eqnarray}
E_a\ni  R(M)^2\quad \leftarrow \quad \varphi\ni  R(M)R(Q)\quad \leftarrow L_{ij}\ni R(Q)^2.
\end{eqnarray}
Thus, $L^{ij}$ is of Weyl-weight 3 and an $SU(2)_R$ triplet,
constructed solely from the Weyl multiplet. 
Hence $L^{ij}$ should be given by  
\begin{eqnarray}
 L^{ij}[{\XW}^2]&=&i\bar\hatR_{ab}{}^{(i}(Q)\hatR^{ab j)}(Q)
+A_1i\bar\chi ^{(i}\chi ^{j)}+A_2v^{ab}\hatR_{ab}{}^{ij}(U)
\end{eqnarray}
for suitable coefficients $A_{1,2}$.
This quantity must be invariant under $S$ transformations
to be the lowest component of a linear multiplet. 
The transformation
\begin{eqnarray}
 \delta _S(\eta )L^{ij}[{\XW}^2]&=&8i\bar \eta ^{(i}\hatR_{ab}^{j)}(Q)v^{ab}
-8i\bar \eta ^{(i}\gamma _{ab}v^{ab}\chi ^{j)}A_1\nn
&&{}+ \left(6i\bar \eta ^{(i}\hatR_{ab}^{j)}(Q)v^{ab}
-\myfrac i2\bar\eta ^{(i}\gamma _{ab}\chi ^{j)}\right)A_2
\end{eqnarray}
fixes $A_2=-4/3$ and $A_1=1/12$.
Then, the embedding formula
is determined by a straightforward but tedious and lengthy
repeated application of the supersymmetry transformation: 
\begin{eqnarray}
  L^{ij}[{\XW}^2]&=&i\bar\hatR_{ab}{}^{(i}(Q)\hatR^{ab j)}(Q)
+\myfrac1{12}i\bar\chi ^{(i}\chi ^{j)}-\myfrac43v^{ab}\hatR_{ab}{}^{ij}(U),\nn
\varphi ^i[{\XW}^2]&=&
\myfrac1{12}\chi ^iD+\myfrac14\gamma _{ab}\hatR_{cd}{}^i(Q)\hatR^{abcd}(M)
-\hatR^{abi}{}_j(U)\left(\hatR_{ab}^j(Q)+\myfrac1{12}\gamma _{ab}\chi ^j\right)\nn
&&{}+8\gamma _{[c}\hatD^c\hatR_{a]b}{}^i(Q)v^{ab}
-2\gamma _c\hatR_{ab}{}^i(Q)\hatD^av^{bc}\nn
&&{}-\myfrac13\gamma _{[a}\hatD_{b]}\chi ^iv^{ab}
+\myfrac16\gamma ^{ab}\gamma ^c\chi ^i\hatD_av_{bc}
-\myfrac23\gamma _{ab}\hatR_{cd}{}^i(Q)v^{ac}v^{bd},\nn
E_a[{\XW}^2]&=&
-\myfrac18\epsilon _{abcde}\hatR^{bcfg}(M)\hatR^{de}{}_{fg}(M)
+\myfrac1{6} \epsilon_{abcde}
\hatR^{bcij}(U)\hatR^{de}{}_{ij}(U)\nn
&&{}
+\hatD^b\left(-\myfrac23 v_{ab} D + 2\hatR_{abcd}(M)v^{cd}
-\myfrac83\epsilon_{abcde}v^{cf}\hatD_fv^{de} \right. \nn
&&\left. {}-4\epsilon_{abcde}v^c{}_f\hatD^dv^{ef} 
+ \myfrac{16}3 v_{ac} v^{cd} v_{db} + \myfrac43 v_{ab} v^2\right),\nn
N[{\XW}^2]&=&\myfrac16D^2+\myfrac14\hatR^{abcd}(M)\hatR_{abcd}(M)
-\myfrac23\hatR_{abij}(U)\hatR^{abij}(U)\nn
&&{}-\myfrac{2}3\hatR_{abcd}(M)v^{ab}v^{cd}
+\myfrac{16}3v_{ab}\hatD^b\hatD_cv^{ac}
+\myfrac83\hatD^av^{bc}\hatD_av_{bc}
+\myfrac83\hatD^av^{bc}\hatD_bv_{ca}\nn
&&{}-\myfrac43\epsilon_{abcde}v^{ab}v^{cd}\hatD_fv^{ef}
+8 v_{ab} v^{bc} v_{cd} v^{da} - 2(v_{ab} v^{ab})^2.
\label{eq:embed2}
\end{eqnarray}
Here, we have omitted the terms trilinear in fermions in the expression of
$\varphi^i$ and  the terms including fermions in the expressions 
of $E_a$ and $N$. 
The first non-trivial check comes from the constraint \eqref{eq:Con.E}, 
indicating that the divergence of $E_a$ vanishes. This can hold because 
the divergence of the first line in $E_a$ vanishes, 
by the Bianchi identity,
while the second and third term vanish if we use the identity 
$\hatD^a \hatD^b A_{ab} = 0$ for a $\XK$-invariant, $SU(2)$-singlet, 
antisymmetric tensor $A_{ab}$. Another 
non-trivial check is the $\XK$-invariance of $E_a$ and $N$, and we can see 
that $E_a$ and $N$ are invariant under $\XK$ transformations.

We form an invariant action for off-shell conformal supergravity from the 
linear multiplet constructed above, using the $V\cdot L$ formula. 
The bosonic term is 
\begin{eqnarray}
\cL(\XV \cdot \XL[\XW^2])\big|_{\text{bosonic}}&=&
c_IY^I_{ij}L^{ij}[\XW^2]-\myfrac12 c_I W^I_a E^a[\XW^2] +\myfrac12 c_I M^I N[\XW^2]\nn
&=&-\myfrac43 c_I Y^I_{ij} v^{ab}\hat R_{ab}^{ij}(U)\nn
&&+ \myfrac{1}{16}\epsilon _{abcde}
c_I W^{aI} \hatR^{bcfg}(M)\hatR^{de}{}_{fg}(M)
- \myfrac1{12}\epsilon_{abcde} 
c_I W^{aI} \hatR^{bc}{}_{jk}(U) \hatR^{dejk}(U)\nn
&&{}+\myfrac18 c_IM^I \hatR^{abcd}(M)\hatR_{abcd}(M) 
-\myfrac13 c_IM^I\hatR_{abjk}(U) \hatR^{abjk}(U)\nn
&&{}+ \myfrac1{12} c_I M^I D^2 + \myfrac16 c_I \hat F^{Iab} v_{ab} D 
- \myfrac13 c_I M^I \hatR_{abcd}(M) v^{ab} v^{cd} \nn
&&{}- \myfrac12 c_I \hat F^{Iab}\hatR_{abcd}(M)v^{cd} 
+ \myfrac83 c_I M^I v_{ab} \hatD^b \hatD_c v^{ac}
+ \myfrac43 c_I M^I \hatD^a v^{bc} \hatD_a v_{bc} \nn
&&{}+ \myfrac43 c_I M^I \hatD^a v^{bc} \hatD_b v_{ca} 
- \myfrac23 c_I M^I \epsilon_{abcde} v^{ab} v^{cd} \hatD_f v^{ef}
+ \myfrac23 c_I \hat F^{Iab} \epsilon_{abcde}v^{cf}\hatD_fv^{de} \nn
&&{}+ c_I \hat F^{Iab}\epsilon_{abcde}v^c{}_f\hatD^dv^{ef} 
- \myfrac43 c_I \hat F^{Iab} v_{ac} v^{cd} v_{db} 
- \myfrac13 c_I \hat F^{Iab} v_{ab} v_{cd} v^{cd} \nn
&&{}+ 4 c_I M^I v_{ab} v^{bc} v_{cd} v^{da}
- c_I M^I (v_{ab} v^{ab})^2
\label{eq:VWW}
\end{eqnarray}
for constants $c_I$. Note that the term containing the second-order 
supercovariant derivative of $v$ depends on the Ricci tensor through 
the $\XK$-gauge field given in \eqref{eq:5DRconstraints}, 
because $\hatD_a v_{bc}$ includes the terms $\sim b_a v_{bc}$ and 
$\sim \omega_{a[b}{}^d v_{c]d}$, and the supercovariant derivative of $b_a$ 
and $\omega_a{}^{bc}$ yields $f_{ab}$\,[ 
see \eqref{eq:dependent-transformations},
\eqref{eq:supercovariant-derivatives}
and \eqref{eq:supercovariant-curvatures}].
The result is
\begin{eqnarray}
v_{ab} \hatD^b \hatD_c v^{ac} 
= v_{ab} \calD^b \calD_c v^{ac} - \myfrac23 v^{ac} v_{cb} R_a{}^b
   -\myfrac1{12}  v_{ab}v^{ab} R
\end{eqnarray} modulo terms including fermions.

\subsection{On-shell Poincar\'e supergravity}

We consider $c_I$ to represent a small perturbation and
add the resulting formula 
$\cL_1\equiv \cL(\XV \cdot \XL[\XW^2])$, 
\eqref{eq:VWW}, to the zeroth-order terms
$\cL_0=\cL_H-\myfrac12\cL_V$, fix the gauge
$\cA^\alpha_i=\delta^\alpha_i$, and eliminate auxiliary fields.
For the components in the Weyl multiplet, the 
equations of motion including fermionic fields can be obtained as follows. 
First, from the analysis of the Weyl-weight, we can see that the omitted terms in 
\eqref{eq:embed2} are independent of the auxiliary field $D$. Thus, the full 
equation of motion for $D$ can be computed, and it is given by
\begin{eqnarray}
\myfrac34 (\cA^2 + 2 \cN) + c_I (M^I D + F_{ab}^I v^{ab} 
+ i \bar{\Omega}^I \chi) = 0.\label{eq:Deom}
\end{eqnarray}
Then, by taking superconformal transformations of this equation, we obtain 
the full equations of motion for other components in the Weyl multiplet.

The first-order correction to the Lagrangian
is obtained by substituting the zeroth-order solution for the 
auxiliary fields \eqref{eq:VWmixing} and \eqref{eq:aux-sol} 
into \eqref{eq:VWW}. Note that 
the appearance  of the $D^2$ term in (\ref{eq:VWW}) changes 
the role of $D$, as can be seen from \eqref{eq:Deom}.
That is, it is no longer a Lagrange multiplier
enforcing the constraint ${\cal N}=1$, 
but, instead, it gives a steep potential $({\cal N}-1)^2/(c_IM^I)$ 
minimized at ${\cal N}=1$.
  Another important point is 
that, from \eqref{eq:VWmixing}, the  gauge field $V^{ij}_\mu$
for the generator $\XU$ is identified with a suitable combination
of the gauge fields $W^I_\mu$ for the gauged supergravity.  

Thus, the supersymmetric completion 
$\cL_1$  of the $W\wedge \tr R\wedge R$ term 
in the on-shell Poincar\'e supergravity becomes 
\begin{eqnarray}
\cL_1=&&{} \epsilon ^{abcde}\left(
\myfrac1{16}c_I W^I_a R_{bc}{}^{fg}R_{defg} -
\myfrac1{6} c_I P_J P_K W^I_a \hat F_{bc}^J \hat F_{de}^K\right)\nn
&&{}+\myfrac18 c_IM^I 
\left(R^{abcd}R_{abcd}-\myfrac43R_{ab}R^{ab}+\myfrac16 R^2 \right)
\nn
&&{}-\myfrac23 c_IP_JP_K M^I\hat F_{ab}^J \hat F^{ab}{}^K
+\myfrac43 \left(c_I (\cN^{-1})^{IJ}P_J\right)
(\cN_I F^I_{ab}) (P_I F^I{}^{ab})
+ \cdots\label{eq:VWWfixed},
\end{eqnarray}
where we have only kept terms necessary for our subsequent analysis. 
Note that we used 
\eqref{eq:R-is-traceless} to express
the supercovariant curvature 
$\hatR_{abcd}(M)$ in terms of  
the metric curvature $R_{abcd}$. For ungauged supergravity ($P_I=0$),
 our result
is consistent with
those obtained by dimensional reduction 
from M-theory\cite{Ferrara1,Ferrara2}.
The important point here is that the supersymmetric completion of  
the $W\wedge \tr R\wedge R$ term in gauged supergravity
not only introduces an $M R^2$
term in the action but also modifies the gauge kinetic and 
gauge Chern-Simons terms.

\section{Condition for the supersymmetric AdS solution}
\label{sec:sugraAdS}
As an application of the $R^2$ term constructed in the previous section,
let us study how it modifies the condition for the supersymmetric AdS
solution.
One merit of the superconformal formalism
presented above is that it allows us to study the supersymmetry
condition in a manner that is largely independent of the action itself.

Let the metric of the AdS space be 
\begin{equation}
 L^2\left( u^2 \eta_{\alpha\beta} dx^\alpha dx^\beta-\frac{du^2}{u^2} \right).
\end{equation} 
where $\alpha,\beta=0,1,2,3$, $\eta=\text{diag}(+,-,-,-)$,
$u=x^4$ and $L$ is the curvature radius.
We further suppose that any field with non-zero spin is zero.
We start with the fact that in such a background, the equation
\begin{equation}
D_\mu \varepsilon - \myfrac{i}{2L} \gamma_\mu \varepsilon=0 
\label{eq:cov-const}
\end{equation}has eight linearly-independent solutions.
Here, $D_{\mu}$ denotes the derivative covariant with respect to local Lorentz 
transformations, and $\varepsilon$ is a spinor without the $SU(2)$-Majorana 
condition. 
If the $i=1$ component of an $SU(2)$-Majorana spinor $\varepsilon^i$
satisfies \eqref{eq:cov-const},
then the $i=2$ component instead satisfies
 \begin{equation}
D_\mu \varepsilon^{i=2} + \myfrac{i}{2L} \gamma_\mu \varepsilon^{i=2}=0.
\end{equation}
Thus, to express it covariantly under $SU(2)_R$, one needs to
introduce a unit three-vector $\vec q$ so that
\begin{equation}
D_\mu \varepsilon^i - \myfrac{1}{2L} \gamma_\mu i
(\vec q\cdot \vec \sigma)^i{}_j\varepsilon^j=0.
\end{equation}
The supersymmetry transformation
of the gravitino \eqref{eq:susytr-weyl} can then be made zero
by choosing  \begin{equation}
\eta^i=\myfrac1{2L}(i\vec q\cdot \vec \sigma)^i{}_j\varepsilon^j.
\end{equation}
The supersymmetric transformation which remains after the gauge fixing
is \begin{equation}
\delta'_Q(\varepsilon)=\delta_Q(\varepsilon)+
\delta_S\left(\myfrac1{2L}(i\vec  q \cdot \vec\sigma)\varepsilon\right).
\label{eq:Qp}
\end{equation}
The vanishing of $\delta_Q\chi^i$  implies that $D=0$.

Next, the vanishing of the gaugino transformation \eqref{eq:susytr-vec}
requires 
\begin{equation}
Y^I{}^i{}_j \varepsilon^j -\myfrac1{2L}M^I (i\vec  q \cdot \vec\sigma)^i{}_j\varepsilon^j=0
\end{equation} for all $I$.  This relation is satisfied for the maximal number
of $\varepsilon^i$ if and only if 
\begin{equation}
Y^I_{ij}= \myfrac1{2L}(i \vec q\cdot \vec\sigma)_{ij} M^I.
\label{eq:AdS-condition}
\end{equation} 
We can set $i \vec q\cdot \vec\sigma = i \sigma^3$ without loss of
generality.
The vanishing of the transformation of the hyperino 
$\delta \zeta^\alpha=0$ under the gauge fixing 
$\cA^\alpha_i\propto \delta^\alpha_i$
determines the curvature radius as 
\begin{eqnarray}
 L=\myfrac32(P_IM^I)^{-1}.
\end{eqnarray}  
Another interesting condition comes from the $[\delta'_Q,\delta'_Q]$
commutator.  From \eqref{eq:QQcommutator}, \eqref{eq:SQcommutator}
and \eqref{eq:QQcommutator2}, it is 
\begin{eqnarray}
[\delta'_Q(\varepsilon),\delta'_Q(\varepsilon')]&=&\delta_U\left(
-\myfrac{6}L \bar\varepsilon^{(i}(i \sigma^3)^{j)}{}_k\varepsilon'{}^{k}\right)
+ \delta_G(-2iM^I \bar\varepsilon\varepsilon' )\nonumber\\
&=&\delta_U\left(
2 P_I M^I (i \sigma^3)^{ij} \bar\varepsilon\varepsilon'\right)
+ \delta_G(-2iM^I \bar\varepsilon\varepsilon' )\nonumber\\
&=&\delta'_G(-2iM^I \bar\varepsilon\varepsilon' ),\label{eq:QpQpcommutator}
\end{eqnarray}
where $\delta_G'$ is the surviving gauge transformation under the
condition $\cA^\alpha{}_i\propto\delta^\alpha_i$ defined in 
(\ref{eq:modG}).
This implies that
 $\delta'_G(M^I)$ should leave the scalar VEVs invariant 
if we consider additional charged matter fields.

  The reader can check that
the analysis up to this point
 does not use any specific property of the action.
Thus it is applicable to any $d=5$ $\cN=2$ supergravity
Lagrangian with arbitrarily complicated higher derivative terms.

Now, let us write down the condition \eqref{eq:AdS-condition}
for our Lagrangian $\cL_0+\cL_1$. To the first order
in $c_I$, $Y^I_{ij}$ is given by the same expression as in \eqref{eq:aux-sol}, 
\begin{equation}
Y^I_{ij}=2 (\cN^{-1})^{IJ}P_J (i \sigma^3)_{ij}.
\end{equation} 
Substituting this into \eqref{eq:AdS-condition}, we obtain
\begin{equation}
P_I=\myfrac14 \cN_{IJ} M^J/L =\myfrac32 
c_{IJK} M^JM^K/L.\label{eq:concrete-AdS-condition}
\end{equation}
This is the attractor equation in 5d gauged supergravity first found in
Ref.\citen{5dgaugedattractor}{}.
By multiplying this equation by $M^I$
we find the condition ${\cal N}=c_{IJK}M^IM^JM^K=1$ again.
One can check that it solves the modified equations of motion which
follows from $\cL_0+\cL_1$. 
The correction to the
potential $({\cal N}-1)^2$ does not shift the VEVs of the scalars,
since the solution before considering higher derivative corrections 
satisfies ${\cal N}=1$, minimizing the added potential.
Note that higher terms with respect to the hatted curvature 
$\hat R_{abcd}(M)$ do not change the AdS solution, since 
the AdS background gives $\hat R_{abcd}(M)=0$.

\section{Comparison to the $a$-maximization}
\label{sec:amax}
\subsection{Brief review of $a$-maximization}
The $a$-maximization is a powerful
technique to uncover the dynamics of
$\cN=1$ superconformal field theories (SCFT) in four dimensions\cite{IW}.
It determines the $R$-symmetry $R_{SC}$ entering into the 
four-dimensional superconformal group as a linear
combination $r^I G_I$ of the $U(1)$ symmetries $G_I$ of the theory.
The AdS dual of the $a$-maximization has been studied\cite{adsamax,currentcorrelator}, and it was found that the
dual is precisely the supersymmetry condition for the AdS solution.
The investigations in Refs.\citen{adsamax} and \citen{currentcorrelator}
are restricted
to the vanishing $U(1)$-gravity-gravity anomaly, corresponding to
the vanishing of the $W\wedge \tr R\wedge R$ contribution.
This is because its supersymmetric completion was not known at that time.
The aim of this section is to extend the analysis of
Refs.\citen{adsamax} and \citen{currentcorrelator}
to the case with non-zero $W\wedge \tr R\wedge R$.

Let us denote by $G_I$, ($I=0,1,\ldots,n_V$)
the conserved $U(1)$ charges of the theory.
$G_I$ also acts on the supercharges.
We denote them by $P_I$:
\begin{equation}
[G_I, Q_\alpha]=\hat P_I Q_\alpha. \label{chargesupercharge}
\end{equation}

The anomaly among global $U(1)$ symmetries
can described through the descent construction using 
the Chern-Simons term in five dimensions,
\begin{equation}
\int \frac{1}{24\pi^2} \hat c_{IJK} W^I\wedge F^J\wedge F^K,\label{gaugeCS}
\end{equation}
where $W^I$ is the gauge field for $G_I$,
and $F^I=F^I_{\mu\nu}dx^\mu\wedge dx^\nu/2$
 is the curvature two-form.
The constants $\hat c_{IJK}$ are given by 
\begin{equation}
\hat c_{IJK}=\tr G_I G_J G_K, \label{c-hat}
\end{equation} where the trace is over the labels of the Weyl fermions of the theory.
It is known that under the AdS/CFT correspondence,
the Chern-Simons interaction \eqref{gaugeCS} is present in the Lagrangian
in five dimensions. Similarly, the $U(1)$-gravity-gravity
anomaly characterized by 
\begin{equation}
\hat c_I=\tr G_I
\end{equation} 
is manifested as the mixed gauge-gravitational Chern-Simons term 
\begin{equation}
 \int \frac{1}{192\pi^2}\hat c_{I}W^I \wedge \tr R \wedge R,\label{gravCS}
\end{equation}where $R$ is the curvature two-form constructed
from the metric.\footnote{The coefficients in \eqref{gaugeCS} and \eqref{gravCS}
are dictated by the index theorem, and they can be found in any textbook
on anomalies (see, e.g., Ref.\citen{AW}).}

The anticommutator of the supertranslation $Q_\alpha$  and
the special superconformal transformation\footnote{
 $Q$ and $S$-supersymmetry here should not be confused
  with $Q$ and $S$-supersymmetry in the superconformal tensor
 formalism in five dimensions. 
 Here $Q$ and $S$ are those of the four-dimensional superconformal group.
 In effect the combination of $Q$ and $S$ in five dimensions
 preserved in the AdS background corresponds to
 both $Q$ and $S$ in four dimensions.}
 $S^\alpha$ 
contains  a particular combination of global symmetries
 : \begin{equation}
\{Q_\alpha, S^\alpha \}\sim  r^I G_I. \label{SCA}
\end{equation} We normalize $r^I$ so that
the charge of the supercharge under $r^I G_I$ be 1,
that is, $r^I \hat P_I=1$.
We denote  the superconformal R-symmetry by $R_{SC}=r^I G_I$.

$R_{SC}$ can be used to study various physical properties
of the theory under consideration.  The relation we need 
is that involving the central charges of the theory.
In four dimensions, there are two of them, $a$ and $c$,
which can be expressed in terms of the superconformal R-symmetry 
as follows:
\begin{equation}
a=\frac3{32}(3 \tr R_{SC}^3-\tr R_{SC}),\qquad
c=\frac1{32}(9\tr R_{SC}^3-5\tr R_{SC}).\label{fieldtheoryresults}
\end{equation}

The basic problem here 
is the identification of the superconformal 
R-symmetry 
$R_{SC}=r^I G_I$,
which can be done with the $a$-maximization.\cite{IW}
Let $Q_F=t^I G_I$ be a global symmetry which commutes with
the supercharges, i.e.
$t^I\hat P_I=0.$
The triangle diagram with one $Q_F$ and two $R_{SC}$ insertions
can be mapped, using the superconformal transformation,
to the triangle diagram with $Q_F$ and two energy-momentum tensor
insertions. The coefficient was calculated precisely and yields the relation 
\begin{equation}
	9\tr Q_F R_{SC} R_{SC} = \tr Q_F.\label{amax1}
\end{equation} 
Another requirement is that 
$\tr Q_F Q_F R_{SC}$ be negative definite.
This comes from the positivity of the 
two-point function of the currents.
Let us introduce
the trial $a$-function $a(s)$ for a trial R-charge $R(s)=s^\Lambda Q_\Lambda$
by generalizing \eqref{fieldtheoryresults}:
\begin{equation}
a(s)=\frac3{32}(3\tr R(s)^3 -\tr R(s) ).
\end{equation}
The conditions in \eqref{amax1}
imply that $a(s)$ is locally maximized 
under the condition $P_I s^I=1$,
at the point $s^I = r^I$, which gives $R_{SC}$.
This is the origin of the terminology of $a$-maximization.

For our purposes, it is convenient to rewrite $a(s)$ as \begin{equation}
a(s)=\myfrac{3}{32}(3\hat c_{IJK}s^Is^Js^K - \hat c_I s^I)
=\myfrac{3}{32}(3\hat c_{IJK}-\hat c_I \hat P_J \hat P_K )s^Is^Js^K,
\label{trick}
\end{equation} where we have used $\hat P_I s^I=1$.

\subsection{Analysis in the AdS space}
Let us suppose that the dual theory in the AdS has the Lagrangian
$\cL_0+\cL_1$.  
We now re-derive $a$-maximization using the 
AdS/CFT prescription and the supergravity analysis
presented in \S\ref{sec:sugraAdS}.

First, \eqref{eq:VWmixing} shows that the gravitino has charge $\pm P_I$
with respect to the gauge fields $W^I_\mu$.
This fact is translated into 
$[G_I, Q_\alpha]=P_I Q_\alpha$ under the AdS/CFT duality,
which allows us to identify $P_I$ with $\hat P_I$, which was
defined in \eqref{chargesupercharge}.

Next, by comparing the gauge Chern-Simons term \eqref{gaugeCS} 
corresponding to the anomaly and the gauge Chern-Simons term
in our Lagrangian $\cL_0+\cL_1$, we get \begin{equation}
\hat c_{IJK}=12\pi^2 c_{IJK}-16\pi^2 c_{(I}P_JP_{K)},
\end{equation}whereas the comparison of \eqref{gravCS}
and the gravitational Chern-Simons term in $\cL_0+\cL_1$ yields
\begin{equation}
\hat c_I = -48\pi^2 c_I.
\end{equation}
Thus we see that 
$c_{IJK}$ entering the Lagrangian is given by \begin{equation}
c_{IJK}=\frac1{12\pi^2} (\hat c_{IJK}
- \myfrac13 \hat c_{(I} P_J P_{K)}).\label{cijk}
\end{equation}

Then, the supersymmetry condition for the AdS space is given by 
\eqref{eq:concrete-AdS-condition}, \begin{equation}
c_{IJK} \vev{M^J} \vev{M^K} \propto P_I,
\end{equation}
where we indicated the scalar VEV at the AdS solution by enclosing in brackets.
Using a Lagrange multiplier, 
the condition above is equivalent to
\begin{equation}
\text{the extremization
of $P_I M^I$ under the constraint $c_{IJK}M^IM^JM^K=1$. }
\end{equation}
Let us define $t^I=M^I/P_I M^I$.  Then this can be  further 
translated as
\begin{equation}
\text{the extremization of $c_{IJK}t^It^Jt^K$ under the condition $P_I t^I=1$.}
\end{equation}
The important point here is that $
a(t)=\myfrac{27\pi^2}{8}c_{IJK}t^It^Jt^K.
$  Thus, we have found that the supersymmetry
condition for the AdS space is given by the extremization of 
the same function $a(t)$.

Finally, we would like to relate the value $\vev{t^I}=\vev{M^I}/(P_I\vev{M^I})$
at the extrema and $r^I$ entering into $R_{SC}=r^IG_I$.
The 4d supercharge corresponds to  $\delta'_Q$
defined in \eqref{eq:Qp}.  Thus, $R_{SC}$ should be a 
linear combination of $U(1)$ generators in the $[\delta'_Q,\delta'_Q]$
commutator
\eqref{eq:QpQpcommutator}, which implies that 
 $r^I\propto \vev{M^I}$.  From the normalization of $r^I$, we should have
$ r^I=\vev{t^I}$. Thus, we find that 
\begin{equation}
\text{$r^I$ is the value of $s^I$ which extremizes $a(s)$ under 
the condition $P_I s^I=1$,}
\end{equation}which is precisely the statement of 
the $a$-maximization procedure. One can also check that the condition
$\tr Q_F Q_F R_{SC}<0$ is equivalent to the positivity of the metric
of the scalar fields at $M^I=\vev{M^I}$, just as is the case with 
$c_I=0$, analyzed in Refs.\citen{adsamax} and \citen{currentcorrelator}.

As a final exercise in this paper, let us calculate the central
charge $a$ from the bulk AdS perspective. 
The method to obtain the central charge $a$ and $c$
in higher derivative gravity theory was pioneered in
Refs.\citen{ON} and \citen{GN}.
and was extended to the general Lagrangian 
in Ref.\citen{FMS}.  In the latter paper,  the formula for the central charge $a$
and $c$
for the boundary CFT with the bulk Lagrangian \begin{equation}
e^{-1}\cL=\myfrac12\left(\frac{12}{L^2}-
\frac{80\alpha+16\beta+8\gamma}{L^4}-R
+\alpha R^2+\beta R_{\mu\nu}^2+\gamma R_{\mu\nu\rho\sigma}^2\right)
\label{general-higher}
\end{equation} is determined to be \begin{eqnarray}
a&=&\pi^2 L^3 \left(1-{40}\alpha- 8\beta-4\gamma\right),\nn
c&=&\pi^2L^3  \left(1-{40}\alpha- 8\beta+4\gamma\right),
\end{eqnarray} where the parametrization of the cosmological
constant in \eqref{general-higher} is chosen so that
the resulting AdS space has curvature radius $L$.

Our Lagrangian $\cL_0+\cL_1$ corresponds to the case \begin{equation}
(\alpha,\beta,\gamma)=\myfrac14{c_IM^I}(\myfrac16,-\myfrac43,1).
\end{equation}Thus we obtain \begin{equation}
a=\pi^2 L^3=\myfrac{27}8\pi^2 (M_IP^I)^{-3}=
\myfrac{27}{8}\pi^2 c_{IJK}t^It^Jt^K=
\myfrac 3{32}\left(3\hat c_{IJK}t^It^Jt^K-\hat c_I t^I  \right),
\end{equation}which is the identical result obtained in
\eqref{fieldtheoryresults}.

\section{Summary and discussion}
\label{sec:conclusion}

In this paper, we have seen how the superconformal formalism
can be used to construct the supersymmetric completion
of the mixed gravitational Chern-Simons term, $c_IW^I\wedge \tr R\wedge R$,
in 5d $\cN=2$ gauged supergravity.
In addition to the known term $c_I M^I R^2$,
we have identified a new contribution in the supersymmetric completion,
namely the modification to the gauge kinetic and Chern-Simons terms.

We also analyzed how the BPS equation for the maximally supersymmetric
AdS solution is modified by the supersymmetric higher derivative term 
constructed above. It was shown that it correctly reproduces
the $a$-maximization of the boundary CFT in the case that the 
mixed $U(1)$-gravity-gravity anomaly exists.

Regarding the outlook for future studies, 
there is a great opportunity for research
using our new $R^2$ term in supergravity.
As discussed in the introduction, such a term naturally arises
when one compactifies string theory down to five dimensions.
Thus, it would be interesting to see, for example, how these terms 
affect the entropy of the five-dimensional black rings and black holes.

Another interesting problem would be to study the effects of
these terms on the  
many exact supersymmetric solutions
to five-dimensional supergravity that were recently derived.
They were found by exploiting the BPS equation fully.
As we now have the full supersymmetry transformation law
including the higher derivative correction,
we should be able to extend their analysis to our case.
We hope to revisit these problems in the future.

\section*{Acknowledgements}
YT would like to acknowledge various helpful discussions with
M. G\"unaydin and  S. Matsuura.
He would also like to express his sincere gratitude to 
the members of the Aspen Center for Physics, where most of this work was
done.
The authors would like to thank a referee for insightful comments
which greatly helped to improve this paper.

The work of KO is supported by the Japan Society for the Promotion
of Science (JSPS) under the Post-doctoral Research Program.
YT was partially supported 
by JSPS Research Fellowships
for Young Scientists when the authors started this work. 
He is now supported by the United States DOE Grant DE-FG02-90ER40542.
\appendix

\section{Notation}
\label{sec:notation}

We summarize our notational conventions in this appendix. 
Firstly, the components of various multiplets and
their basic properties
are summarized in Table \ref{table}.
The gamma matrices $\gamma^a$ satisfy
$\{\gamma^a,\gamma^b\}=2\eta^{ab}$ and $(\gamma^a)^{\dagger}=\eta_{ab}\gamma^b$,
where $\eta^{ab}={\rm diag}(+,-,-,-,-)$.
\,$\gamma_{a\ldots b}$ represents an antisymmetrized product
of gamma matrices:
\begin{equation}
\gamma_{a\ldots b}=\gamma_{[a}\ldots\gamma_{b]},
\end{equation}
where the square brackets denote complete antisymmetrization
with weight 1. Similarly $(\ldots)$ denote complete
symmetrization with weight 1.
We choose the Dirac matrices to satisfy
\begin{equation}
\gamma^{a_1\ldots a_5}=\epsilon^{a_1\ldots a_5},
\end{equation}
where $\epsilon^{a_1\ldots a_5}$ is a totally antisymmetric
tensor with $\epsilon^{01234}=1$.

The $SU(2)$ index $i$ ($i$=1,2) is raised and lowered with $\epsilon_{ij}$,
where $\epsilon_{12}=\epsilon^{12}=1$, in the
northwest-southeast (NW-SE) convention:
\begin{equation}
A^i=\epsilon^{ij}A_j,\hspace{2em}A_i=A^j\epsilon_{ji}.
\end{equation}

The charge conjugation matrix C in 5D has the properties
\begin{equation}
C^T=-C,\hspace{2em}C^{\dagger}C=1,\hspace{2em}C\gamma_aC^{-1}=\gamma_a^T.
\end{equation}
Our five-dimensional spinors satisfy the $SU(2)$-Majorana condition
\begin{equation}
\bar\psi^i\equiv\psi_i^{\dagger}\gamma^0=\psi^{iT}C,
\end{equation}
where the spinor indices are omitted. When the $SU(2)$ indices are suppressed
in the bilinear terms of spinors, the NW-SE contraction is understood, e.g.
$\bar\psi\gamma^{a_1\ldots a_n}\lambda=\bar\psi^i\gamma^{a_1\ldots a_n}\lambda_i$.
Changing the order of the spinors in a bilinear leads to the following
signs:
\begin{equation}
\bar\psi\gamma^{a_1\ldots a_n}\lambda=
(-1)^{(n+1)(n+2)/2}\bar\lambda\gamma^{a_1\ldots a_n}\psi.
\end{equation} 
If the $SU(2)$ indices are not contracted, the sign switches.
We often use the Fierz identity, which in 5D reads
\begin{equation}
\psi^i\bar\lambda^j=-\frac{1}{4}(\bar\lambda^j\psi^i)
-\frac{1}{4}(\bar\lambda^j\gamma^a\psi^i)\gamma_a
+\frac{1}{8}(\bar\lambda^j\gamma^{ab}\psi^i)\gamma_{ab}.
\end{equation}

\begin{table}[12p,t]
\caption{Multiplets in 5D superconformal gravity.}
\label{table}
\begin{center}
\begin{tabular}{ccccc} \hline \hline
    field      & type   & remarks & $SU(2)$ & Weyl-weight    \\ \hline 
\Tspan{$e_\mu{}^a$} &   boson    & f\"unfbein    & \bf{1}    &  $ -1$   \\  
$\psi^i_\mu$  &  fermion  & $SU(2)$-Majorana & \bf{2}
&$-\myfrac12$ \\  
\Tspan{$b_\mu$} & boson &  real & \bf{1} & 0 \\
\Tspan{$V^{ij}_\mu$}    &  boson    
& $V_\mu^{ij}=V_\mu^{ji}=(V_{\mu ij})^*$ & \bf{3}&0\\ 
$v_{ab}$&boson& real, antisymmetric &\bf{1}&1 \\
$ \chi^i$  &  fermion  & $SU(2)$-Majorana & \bf{2}    
&\myfrac32 \\  
$D$    &  boson    & real & \bf{1} & 2 \\ \hline
\multicolumn{5}{c}{dependent gauge fields} \\ \hline
$\omega_\mu{}^{ab}$ &   boson    & spin connection & \bf{1}    &   0\\
$\phi_\mu^i$ & fermion & $SU(2)$-Majorana & \bf{2} & \myfrac12\\
$f_\mu{}^a$ & boson & real & \bf{1} & 1\\
\hline 
\multicolumn{5}{c}{Vector multiplet} \\ \hline
$W_\mu$      &  boson    & real gauge field   &  \bf{1}    &   0     \\
$M$& boson & real& \bf{1} & 1 \\ 
$\Omega ^i$& fermion  &$SU(2)$-Majorana  & \bf{2} &\myfrac32 \\  
$Y_{ij}$    &  boson    & $Y^{ij}=Y^{ji}=(Y_{ij})^*$   & \bf{3} & 2 \\ \hline
\multicolumn{5}{c}{Hypermultiplet} \\ \hline
\Tspan{$\calA_i^\alpha$}     &  boson & 
$\calA^i_\alpha=\epsilon^{ij}\calA_j^\beta\rho_{\beta\alpha}=-(\calA_i^\alpha)^*$ &\bf{2}& \myfrac32\\  
$\zeta^\alpha$    &  fermion  & $\bar\zeta^\alpha\equiv(\zeta_\alpha)^\dagger\gamma_0 = \zeta^{\alpha\T}C$ 
& \bf{1}  & 2 \\ 
${\cal F}_i^\alpha$  &  boson    & 
${\cal F}^i_\alpha=-({\cal F}_i^\alpha)^*$  &  \bf{2}   &\myfrac52 \\ \hline
\multicolumn{5}{c}{Linear multiplet} \\ \hline
\Tspan{$L^{ij}$}& boson & $L^{ij}=L^{ji}=(L_{ij})^*$  &  \bf{3}   & 3 \\ 
$\varphi^i$ &  fermion  & $SU(2)$-Majorana & \bf{2}&\myfrac72 \\  
$E_a$ & boson & real, constrained by (\ref{eq:Con.E})& \bf{1} & 4 \\ 
$N$ & boson & real & \bf{1} & 4 \\ \hline
\end{tabular}
\end{center}
\end{table}

\section{Definitions and Useful Formulae for the Weyl Multiplet}
\label{sec:weylformulae}

In this appendix,
we summarize useful formulae for the Weyl multiplet. 
Firstly, the solution to the constraints \eqref{conventional-constraints}
is given by the following:
\begin{eqnarray}
&&\omega _\mu {}^{ab}=\omega ^0_\mu {}^{ab}
+i(2\bar\psi _\mu \gamma ^{[a}\psi ^{b]}+\bar\psi ^a\gamma _\mu \psi ^b)-2e_\mu {}^{[a}b^{b]},\nn
&& \quad \hbox{with}\ \ 
\omega ^0_\mu {}^{ab}\equiv  -2e^{\nu [a}\partial _{[\mu }e_{\nu ]}{}^{b]}
+e^{\rho [a}e^{b]\sigma }e_\mu {}^c\partial _\rho e_{\sigma c}, \nn
&&\phi _\mu ^i=\left(-\myfrac13e^a_\mu \gamma ^b+\myfrac1{24}\gamma _\mu \gamma ^{ab}\right)
\hatR_{ab}^{\prime\, i}(Q),\nn
&&f_\mu {}^a=
\left(\myfrac16\delta _\mu ^\nu \delta ^a_b
-\myfrac1{48}e_\mu ^ae^\nu _b\right)\hatR'_\nu {}^b(M).
\label{eq:5DRconstraints}
\end{eqnarray}
Here $\hatR_\mu {}^a(M)\equiv  \hatR_{\mu \nu }{}^{ba}(M)e^\nu _b$, and 
the primes on the curvatures indicate that 
$\hatR_{ab}^{\prime\,i}(Q)=\hatR_{ab}^i(Q)|_{\phi _\mu =0}$ and 
$\hatR'_\mu {}^a(M)=\hatR_\mu {}^a(M)|_{f_\nu {}^b=0}$. 
The transformation laws of their dependent gauge fields can be
obtained by using those of the independent fields of the Weyl multiplet, 
in principle. 
The explicit $\XK$-transformation laws of the gauge field 
$\omega_{\mu}{}^{ab}$,
\begin{eqnarray}
\delta_K(\xi_K^a) \omega_{\mu}{}^{ab}= - 4 \xi_K^{[a} e_{\mu}{}^{b]},
\label{eq:dependent-transformations}
\end{eqnarray}
are needed to check the $\XK$-invariance of the embedding formulae 
in (\ref{eq:embed2}).

We used two types of covariant derivatives in the main text. The 
first one is the `unhatted' derivative $\calD_\mu $, which is covariant only 
with respect to the homogeneous transformations 
$\XM_{ab},\XD$ and $\XU^{ij}$ 
and the $\XG$ transformation for non-singlet fields under the Yang-Mills 
group $G$. The other is the `hatted' derivative $\hatD_\mu $, which 
denotes the fully superconformal covariant derivative. 
With $h_\mu ^A$ denoting the gauge fields of the transformation $\XX_A$, they 
are defined as
\begin{equation}
\calD_\mu \equiv  \partial _\mu -\!\sum_{\XX_A=\XM_{ab},\XD,\,\XU^{ij}(,\XG)}
\!\!\!h_\mu ^A\XX_A, 
\qquad  
\hatD_\mu = \calD_\mu -\!\sum_{\XX_A=\XQ^i,\XS^i,\XK_a}\!\!h_\mu ^A\XX_A.
\end{equation} 
Below we give the explicit forms of
the covariant derivatives appearing in Eq.(\ref{eq:susytr-weyl}) 
 for convenience: 
\begin{eqnarray}
\calD_\mu \varepsilon ^i&=&\left(\partial _\mu -\myfrac14\omega _\mu {}^{ab}\gamma _{ab}+\half b_\mu \right)
\varepsilon ^i-V_\mu ^{\ i}{}_j\varepsilon ^j, \nn
\calD_\mu \eta ^i&=&\left(\partial _\mu -\myfrac14\omega _\mu {}^{ab}\gamma _{ab}-\myfrac12b_\mu 
\right)\eta ^i-V_\mu ^{\ i}{}_j\eta ^j,\nn
\calD_\mu \xi ^a_K&=&\left(\partial _\mu -b_\mu \right)\xi ^a_K-\omega _\mu {}^{ab}\xi _{Kb}, \nn
\hatD_\mu v_{ab}&=& \partial _\mu v_{ab}
+2\omega _\mu {}_{[a}{}^c v_{b]c}-b_\mu v_{ab}  
+\myfrac{i}8\bar\psi _\mu \gamma _{ab}\chi 
+\myfrac32 i\bar\psi _\mu \hatR_{ab}(Q), \nn
\hatD_\mu \chi ^i&=&\calD_\mu \chi ^i-D\psi _\mu ^i+2\gamma ^c\gamma ^{ab}\psi _\mu ^i\hatD_av_{bc}
-\gamma \dt \hatR(U)^i{}_j\psi _\mu ^j
+2\gamma ^a\psi _\mu ^i\epsilon_{abcde}v^{bc}v^{de}-4\gamma \dt v\phi _\mu ^i,\nn
\calD_\mu \chi ^i&=&\left(\partial _\mu -\myfrac14\omega _\mu {}^{ab}\gamma _{ab}
-\myfrac32b_\mu \right)\chi ^i-V_\mu ^{\ i}{}_j\chi ^j\,.
\label{eq:supercovariant-derivatives}
\end{eqnarray}

The superconformally covariant curvatures $\hatR_{\mu\nu}{}^A$ are 
defined as the commutator of the covariant derivatives:
\begin{eqnarray}
[\hatD_a, \hatD_b] = - \sum_{A=\XQ^i,\XM_{ab},\XD,\XU_{ij},\XS^i,\XK_a}
\hatR_{ab}{}^A \XX_A
\end{eqnarray}
They are given explicitly by the following expressions:
\begin{eqnarray}
\hatR_{\mu\nu}{}^a(P)&=&2\partial_{[\mu}e_{\nu]}{}^a-2\omega_{[\mu}{}^{ab}e_{\nu]b}
+2b_{[\mu}e_{\nu]}{}^a+2i\bar\psi_\mu\gamma^a\psi_\nu,\nn
\hatR_{\mu\nu}{}^i(Q)&=&
2\partial^{\Span{i}}_{[\mu}\psi_{\nu]}^i-\myfrac12\omega_{[\mu}{}^{ab}\gamma_{ab}\psi_{\nu]}^i
+b_{[\mu}\psi_{\nu]}^i-2V_{[\mu}^i{}_j\psi_{\nu]}^j+\gamma_{ab[\mu}\psi_{\nu]}v^{ab}
-2\gamma_{[\mu}\phi_{\nu]}^i,\nn
\hatR_{\mu\nu}{}^{ab}(M)&=&
2\partial_{[\mu}\omega_{\nu]}-2\omega_{[\mu}{}^a{}_c\omega_{\nu]}{}^{cb}
-4i\bar\psi_{[\mu}\gamma^{ab}\phi_{\nu]}
+2i\bar\psi_{[\mu}\gamma^{abcd}\psi_{\nu]}v_{cd}\nn
&&+4i\bar\psi_{[\mu}\gamma^{[a}\hatR_{\nu]}{}^{b]}(Q)
+2i\bar\psi_{[\mu}\gamma_{\nu]}\hatR^{ab}(Q)
+8f_{[\mu}{}^{[a}e_{\nu]}{}^{b]},\nn
\hatR_{\mu\nu}(D)&=&2\partial_{[\mu}b_{\nu]}+4i\bar\psi_{[\mu}\phi_{\nu]}+4f_{[\mu\nu]},\nn
\hatR_{\mu\nu}{}^{ij}(U)&=&2\partial^{\Span{j}}_{[\mu}V_{\nu]}^{ij}
-2V_{[\mu}^i{}_kV_{\nu]}^{kj}+12i\bar\psi_{[\mu}^{(i}\phi_{\nu]}^{j)}
-4i\bar\psi_{[\mu}^i\gamma\dt v\psi_{\nu]}^j
+\myfrac{i}2\bar\psi_{[\mu}^{(i}\gamma^{\Span{(i}}_{\nu]}\chi^{j)}_{\Span{[\mu}}\,,\nn
\hatR_{\mu\nu}{}^i(S)&=&2\partial^{\Span{i}}_{[\mu}\phi_{\nu]}^i
-\myfrac12\omega_{[\mu}{}^{ab}\gamma_{ab}\phi_{\nu]}^i
-b_{[\mu}\phi_{\nu]}^i-2V_{[\mu}^i{}_j\phi_{\nu]}^j+2f_{[\mu}^{\ \,a}\gamma_a\psi_{\nu]}^i
+\cdots,\nn
\hatR_{\mu\nu}{}^a(K)&=&2\partial_{[\mu}f_{\nu]}{}^a-2\omega_{[\mu}{}^{ab}f_{\nu]b}
-2b_{[\mu}f_{\nu]}{}^a+2i\bar\phi_\mu\gamma^a\phi_\nu+\cdots\,.\label{eq:supercovariant-curvatures}
\end{eqnarray}
Here, the dots in the $\XS^i$ and $\XK^a$ curvature expressions 
denote terms containing other curvatures. 

To compute the $\XQ$-variation of the covariant derivatives of 
some fields, the following formula is useful:
\begin{equation}
[\delta_Q, \hatD_a] = 
- \delta_Q ([\delta_Q \psi_a^i]_{\rm cov})) 
- \delta_S ([\delta_Q \phi_a^i]_{\rm cov})) 
+ \cdots.
\end{equation}
Here the fermionic terms are omitted and $[\cdots]_{cov}$ denotes 
the covariant 
part of the variations, namely, 
\begin{eqnarray}
[\delta_Q \psi_{a}^i]_{\rm cov} &=& \myfrac12 \gamma_{abc} v^{bc} 
\varepsilon^i ,\nn
{}[\delta_Q \phi_{a}^i]_{\rm cov} &=& \myfrac13 \Big( \hatR_{ab}{}^i{}_j(U) 
\gamma^b 
- \myfrac18 \gamma_{a} \gamma \cdot \hatR^i{}_j(U) \Big) \varepsilon^j \nn
&&{}-\myfrac{1}{12}\Big( 3\hatD_{\mu} \gamma \cdot v \varepsilon^i 
+ \gamma_{abcd} \hatD^b v^{cd} \varepsilon^i + \gamma_{ab} \hatD_c v^{cb} 
\varepsilon^i - 2 \gamma^{bc} \varepsilon^i \hatD_b v_{ca} 
- 3 \varepsilon^i \hatD^b v_{ba} \nn
&&{}-\gamma_{abcde} \varepsilon^i v^{bc} v^{de} + 4 v_{ab} v_{cd} \gamma^{bcd} 
\varepsilon^i + 16 v_{ab} v^{bc} \gamma_c \varepsilon^i 
+ 5 v_{bc}v^{bc} \gamma_a \varepsilon^i \Big).
\end{eqnarray}
Using this, we can verify that the variations of the supercovariant curvatures 
 not contain any term non-covariant with respect to the superconformal 
transformations.

Finally, we present the explicit forms of 
the variations of the supercovariant curvatures 
$\hatR^i(Q)$ and $\hatR^{ij}(U)$:
\begin{eqnarray}
\delta \hatR_{ab}^i(Q)&=&
-\myfrac14\hatR_{ab}{}^{cd}(M)\gamma _{cd}\varepsilon ^i
-\myfrac13\hatR_{ab}{}^i{}_j(U)\varepsilon ^j
+\myfrac1{12}\gamma \cdot \hatR^{i}{}_j(U)\gamma _{ab}\varepsilon ^j\nn
&&{}+\myfrac16\gamma _{abcde}\hatD^cv^{de}\varepsilon ^i+
\myfrac13\hatD_{[a}v^{cd}\gamma _{b]cd}\varepsilon ^i+
\myfrac16\gamma _{abc}\hatD_dv^{dc}\varepsilon ^i\nn
&&{}-\myfrac23\hatD_{[a}v_{b]c}\gamma ^c\varepsilon ^i
+\myfrac13\hatD^cv_{c[a}\gamma _{b]}\varepsilon ^i+\myfrac13\slashD v_{ab}\varepsilon ^i\nn
&&{}-\myfrac23v_{ab}\gamma\cdot v\varepsilon ^i
-\myfrac23\gamma_{cd}v_{[a}{}^cv_{b]}{}^d\varepsilon ^i
+\myfrac43\gamma_{[a}{}^cv_{b]}{}^{d}v_{cd}\varepsilon ^i
-\myfrac16\gamma_{ab}v^2\varepsilon ^i+ \cdots\nn
&&{}-\myfrac13\gamma _{abcd}v^{cd}\eta ^i
+\myfrac43\gamma _{[a}{}^cv_{b]c}\eta ^i+2v_{ab}\eta ^i, \label{eq:R(Q)tr} \\
\delta  \hat R_{ab}{}^{ij}(U)&=&-6i\bar\varepsilon ^{(i}\hatR_{ab}{}^{j)}(S)
+4i\bar\varepsilon ^{(i}\gamma \cdot v\hatR_{ab}{}^{j)}(Q)
+\myfrac i2\bar\varepsilon ^{(i}\gamma _{[a}\hatD_{b]}\chi ^{j)},\nn
&&{}-\frac{i}4\bar \varepsilon^{(i}\gamma_{abcd}\chi^{j)}v^{cd}
-\frac{i}2\bar \varepsilon^{(i}\gamma_{c[a}\chi^{j)}v_{b]}{}^c\nn
&&{}+6i\bar\eta ^{(i}\hatR_{ab}^{j)}(Q)
-\myfrac i2\bar\eta ^{(i}\gamma _{ab}\chi ^{j)} .
\end{eqnarray}
The ellipsis in \eqref{eq:R(Q)tr} represents
terms trilinear in fermions in $\delta_Q \hatR(Q)$. 
No term of $\delta_S\hatR(Q)$ is omitted.

\section{Conventions for Vector Multiplets}
\label{sec:vector-convention}
Here, we summarize the conventions for the vector multiplets in the 
original on-shell formalism of G\"unaydin, Sierra and Townsend\cite{GST1,GST2},
in the superconformal formalism of Fujita, Kugo and Ohashi, \cite{KO,FO}
and in the formalism of Bergshoeff et al.,\cite{Berg0,Berg1}
for convenience. 
The multiplets are labeled as follows: \[
\begin{array}[12p]{rccc}
&\text{scalar}&\text{gaugini}&\text{vector}\\
\hline
\text{GST}:&h^I, & \chi_i^I,&  A^I_\mu  \\
\text{FKO}: & M^I, & \Omega_i^I, &W^I_\mu, \\
\text{Berg.}: & \sigma^I, & \psi_i^I, & A^I_\mu.
\end{array} 
\] 
All groups denote the Chern-Simons coefficients by $c_{IJK}$.
The gauge fields are to be identified according to  
\begin{equation}
A_\mu{}^{\text{GST}}=W_\mu{}^{\text{FKO}}=A_{\mu}{}^{\text{Berg.}},
\end{equation}
while the symbols for the Chern-Simons  are related by 
\begin{equation}
c_{IJK}^{\text{GST}}=-\left(\myfrac32\right)^{3/2}c_{IJK}^{\text{FKO}}, \qquad
c_{IJK}^{\text{Berg.}}=3c_{IJK}^{\text{FKO}}.
\end{equation} 
The scalars are related by 
\begin{equation}
h^I=-\sqrt{\myfrac23}M^I, \qquad
\sigma^I=-M^I,
\end{equation}and the very special real manifold is defined, 
respectively, by 
\begin{equation}
c_{IJK}^{\text{FKO}}M^IM^JM^K=1,\qquad
c_{IJK}^{\text{GST}}h^I h^J h^K=1,\qquad
c_{IJK}^{\text{Berg.}}\sigma^I\sigma^J\sigma^K=-3.
\end{equation}


\begin{thebibliography}{99}
    \bibitem{GST1}
  M.~G\"unaydin, G.~Sierra and P.~K.~Townsend,
 Nucl.\ Phys.\ B {\bf 242} (1984), 244.


    \bibitem{GST2}
  M.~G\"unaydin, G.~Sierra and P.~K.~Townsend,
 Nucl.\ Phys.\ B {\bf 253} (1985), 573.

    \bibitem{BR}
  E.~Bergshoeff and M.~Rakowski,
 Phys.\ Lett.\ B {\bf 191} (1987), 399.
 
    \bibitem{MoReview}
  T.~Mohaupt,
 Fortsch.\ Phys.\  {\bf 49} (2001), 3;
  {\tt hep-th/0007195}.
 
    \bibitem{KO}
  T.~Kugo and K.~Ohashi,
 Prog.\ Theor.\ Phys.\ {\bf 104} (2000), 835;
  {\tt hep-ph/0006231}.
 
    \bibitem{FO}
  T.~Fujita and K.~Ohashi,
 Prog.\ Theor.\ Phys.\ {\bf 106} (2001), 221;
  {\tt hep-th/0104130}.
 
    \bibitem{Berg0}
  E.~Bergshoeff, S.~Cucu, M.~Derix, T.~de Wit, R.~Halbersma and A.~Van Proeyen,
 J.High Energy Phys.{\bf 06} (2001), 051;
  {\tt hep-th/0104113}.
 
    \bibitem{Berg1}
  E.~Bergshoeff, S.~Cucu, T.~de Wit, J.~Gheerardyn, S.~Vandoren and A.~Van Proeyen,
 Class.\ Quant.\ Grav.\ {\bf 21} (2004), 3015,
Class.\ Quant.\ Grav.\  {\bf 23} (2006), 7149;
  {\tt hep-th/0403045}.
 
    \bibitem{Ferrara1}
  S.~Ferrara, R.~R.~Khuri and R.~Minasian,
 Phys.\ Lett.\ B {\bf 375} (1996), 81;
  {\tt hep-th/9602102}.
 
    \bibitem{Ferrara2}
  I.~Antoniadis, S.~Ferrara, R.~Minasian and K.~S.~Narain,
 Nucl.\ Phys.\ B {\bf 507} (1997), 571;
  {\tt hep-th/9707013}.
 
    \bibitem{veryspecial}
  B.~de Wit and A.~Van Proeyen,
 Phys.\ Lett.\ B {\bf 293} (1992), 94;
  {\tt hep-th/9207091}.
 
    \bibitem{interpretation-of-higher}
  D.~A.~Eliezer and R.~P.~Woodard,
 Nucl.\ Phys.\ B {\bf 325} (1989), 389.
 
    \bibitem{interpretation-of-higher2}
  T.~C.~Cheng, P.~M.~Ho and M.~C.~Yeh,
 Nucl.\ Phys.\ B {\bf 625} (2002), 151;
  {\tt hep-th/0111160}.
 
    \bibitem{aux-elimination}
  P.~C.~Argyres, A.~M.~Awad, G.~A.~Braun and F.~P.~Esposito,
 J.High Energy Phys.{\bf 07} (2003), 060;
  {\tt hep-th/0306118}.
 
    \bibitem{5dgaugedattractor}
  A.~Chou, R.~Kallosh, J.~Rahmfeld, S.~J.~Rey, M.~Shmakova and W.~K.~Wong,
 Nucl.\ Phys.\ B {\bf 508} (1997), 147;
  {\tt hep-th/9704142}.
 
    \bibitem{IW}
  K.~Intriligator and B.~Wecht,
 Nucl.\ Phys.\ B {\bf 667} (2003), 183;
  {\tt hep-th/0304128}.
 
    \bibitem{adsamax}
  Y.~Tachikawa,
 Nucl.\ Phys.\ B {\bf 733} (2006), 188;
  {\tt hep-th/0507057}.
 
    \bibitem{currentcorrelator}
  E.~Barnes, E.~Gorbatov, K.~Intriligator and J.~Wright,
 Nucl.\ Phys.\ B {\bf 732} (2006), 89;
  {\tt hep-th/0507146}.
 
    \bibitem{AW}
  L.~Alvarez-Gaum\'e and E.~Witten,
 Nucl.\ Phys.\ B {\bf 234} (1984), 269.
 
    \bibitem{ON}
  S.~Nojiri and S.~D.~Odintsov,
 Int.\ J.\ Mod.\ Phys.\ A {\bf 15} (2000), 413;
  {\tt hep-th/9903033}.
 
    \bibitem{GN}
  M.~Blau, K.~S.~Narain and E.~Gava,
 J.High Energy Phys.{\bf 09} (1999), 018;
  {\tt hep-th/9904179}.
 
    \bibitem{FMS}
  M.~Fukuma, S.~Matsuura and T.~Sakai,
 Prog.\ Theor.\ Phys.\ {\bf 105} (2001), 1017;
  {\tt hep-th/0103187}.
 \end{thebibliography}
\end{document}